\documentclass[aps,prl,showpacs,twocolumn,reprint]{revtex4-1}

\usepackage{amsmath, amsthm, amssymb}
\usepackage{graphicx}
\usepackage{bm}

\newcommand{\pa}{\partial}
\newcommand{\tR}{\tau}
\newcommand{\tRB}{\bm \tR}
\newcommand{\tDummy}{t''}
\newcommand{\tSimple}{\tau^{(0)}}
\newcommand{\FI}{\mathcal I}
\newcommand{\FIB}{\bm \FI}
\newcommand{\mcL}{{\mathcal L}}
\newcommand{\fric}{\zeta}
\newcommand{\bfric}{{\bm \fric}}

\newcommand{\tCurr}{t_0}

\newcommand{\springMin}{y}

\newcommand{\cp}{\lambda}
\newcommand{\bcp}{{\boldsymbol\cp}}
\newcommand{\bcpCurr}{\bcp(\tCurr)}

\newcommand{\prot}{\bf \Lambda}
\newcommand{\fConj}{{\bf X}}
\newcommand{\resp}{\chi}
\newcommand{\bResp}{{\boldsymbol\resp}}
\newcommand{\covar}{\Sigma}
\newcommand{\bCovar}{{\boldsymbol\covar}}
\newcommand{\covarCurr}{\covar^{(\bcpCurr)}}
\newcommand{\bCovarCurr}{\bCovar^{(\bcpCurr)}}

\newcommand{\be}{\begin{equation}}
\newcommand{\ee}{\end{equation}}
\newcommand{\ba}{\begin{align}}
\newcommand{\ea}{\end{align}}
\newcommand{\bse}{\begin{subequations}}
\newcommand{\ese}{\end{subequations}}

\newcommand{\md}{d}
\newcommand{\f}{\frac}

\newcommand{\work}{{\mathcal W}}
\newcommand{\exWork}{\work_{\text{ex}}}
\newcommand{\heat}{{\mathcal Q}}
\newcommand{\power}{{\mathcal P}}
\newcommand{\exPower}{\power_{\text{ex}}}
\newcommand{\instD}{\exPower(t_0)}
\newcommand{\kB}{k_{\text{B}}}
\newcommand{\kT}{\kB T}
\newcommand{\la}{\langle}
\newcommand{\ra}{\rangle}
\newcommand{\fricSp}{\fric^{\text{c}}}  
\renewcommand{\l}{\left}
\renewcommand{\r}{\right}

\usepackage[usenames, dvipsnames]{color}
\usepackage[normalem]{ulem}
\newcommand{\das}[1]{#1}

\begin{document}

\title{Thermodynamic metrics and optimal paths}

\author{David A.\ Sivak} 
\email{dasivak@lbl.gov}
\affiliation{Physical Biosciences Division, Lawrence Berkeley National Laboratory, Berkeley, California 94720, USA}

\author{Gavin E.\ Crooks}
\affiliation{Physical Biosciences Division, Lawrence Berkeley National Laboratory, Berkeley, California 94720, USA}

\begin{abstract}
A fundamental problem in modern thermodynamics is how a molecular-scale machine performs useful work, while operating away from thermal equilibrium without excessive dissipation. To this end, we derive a friction tensor that induces a Riemannian manifold on the space of thermodynamic states. Within the linear-response regime, this metric structure controls the dissipation of finite-time transformations, and bestows optimal protocols with many useful properties. We discuss the connection to the existing thermodynamic length formalism, and demonstrate the utility of this metric by solving for optimal control parameter protocols in a simple nonequilibrium model.
\end{abstract}

\date{\today}
\maketitle

\paragraph*{\bf Introduction.}
Molecular machines are microscopic objects that manipulate energy, matter, and information on the nanometer scale. Naturally-occurring machines are central to the performance of virtually any prominent cellular process, and the design of synthetic machines holds out the promise for significant technological advances. A major impediment to quantitative understanding of their thermodynamics is that molecular-scale machines typically operate out of thermodynamic equilibrium. For instance, the rotary $F_OF_1$-ATPase motor is powered by proton flow across a gradient producing a free energy difference of $\sim$200 meV per proton. 
\das{This free energy difference dwarfs} the characteristic energy scale of thermal fluctuations under ambient conditions, $1 \kT\approx 25$ meV (where $\kB$ is Boltzmann's constant and the temperature is $T \sim$ 300 K), \das{hence, the proton flux} drives the \das{machine out of equilibrium}. 
In such contexts equilibrium statistical mechanics has limited applicability and a nonequilibrium understanding of these machines is vital. Indeed, living processes with their preparation and preservation of order must, by their very nature, be out of equilibrium, leading Schr\"{o}dinger to equate death with ``the decay into thermodynamical equilibrium''~\cite{Schrodinger:WhatIsLife}.   

A central figure of merit for both molecular and macroscopic machines is thermodynamic efficiency: their ability to exploit \das{available} energy from a source to perform useful work, while minimizing dissipation of heat into the surrounding environment. The importance of efficiency engenders an interest in understanding the basic physical principles at play, the limits on efficiency in energy conversion, and the characteristics of optimal machines. In order to generate insights into biomolecular machine efficiency, insights that are transferable to the design of novel synthetic molecular machines, a general framework is necessary, one that abstracts away from many of the molecular details and instead focuses (at least initially) on criteria for optimal nonequilibrium processes.   

For macroscopic systems, the properties of optimal processes have been investigated using thermodynamic length, a natural measure of the distance between equilibrium thermodynamic states~\cite{Weinhold:1975wr}.  Its original derivations, developed in the context of finite-time thermodynamics, considered the metrics on equilibrium manifolds, specifically the second derivatives of internal energy~\cite{Weinhold:1975wr}, entropy~\cite{Ruppeiner:1979ve}, or free energy~\cite{Schlogl1985}, all essentially equivalent in the thermodynamic limit~\cite{Salamon:1984wv}. Central results were derived under the assumption of endoreversibility~\cite{Salamon:1983vc}, whereby the system and environment are in thermal equilibrium, though not necessarily equilibrated with each other. In a system driven by changes in a single control parameter, this amounts to the system trailing the environment, at any instant residing in an equilibrium corresponding to the system at a previous value of the control parameter. From these foundations, the second derivatives of internal energy, free energy and entropy were shown to each impose a Riemannian metric structure on the equilibrium surface~\cite{Salamon:1983vc,Brody:1995vx}. 

Our aim is to adapt this framework to microscopic, nanoscale processes rather than the macroscopic processes for which it was originally formulated. Recent extensions have developed a microscopic formulation of thermodynamic length~\cite{Crooks:2007wy} with a metric tensor of Fisher information~\cite{Burbea:1982uo} (equivalent for thermodynamic systems to the equilibrium fluctuations of the conjugate force), and showed how to experimentally measure this quantity using work fluctuation relations~\cite{Feng:2009vf}. In this paper we show that a microscopic and generalized formulation of thermodynamic length analysis can be derived directly from linear-response theory, without recourse to endoreversibility. The resulting thermodynamic metric structure imbues optimal processes with several important properties: optimal paths (those that minimize dissipation) are geodesics~\cite{Nulton:1985vf}, dissipation is inversely proportional to protocol duration, the optimal control parameter path is independent of duration, and optimal protocols perturb the control parameter so as to accumulate excess work at a constant rate~\cite{Salamon:1983vc,Salamon:1985ut}.

\paragraph*{\bf Derivation.} 
Linear response is a standard framework for understanding fluctuations out of equilibrium~\cite{Marconi:2008}. Here we derive a generalized thermodynamic length analysis from linear-response, without resort to the assumption of endoreversibility.  

We assume a physical system in contact with a thermal bath. The probability distribution over microstates $x$ at equilibrium is given by the canonical ensemble
\be
\pi(x|\bcp) = \exp \beta[F(\bcp) - E(x,\bcp)] \ , \label{equ:CanonEns}
\ee
where $\beta = 1/\kT$ is the inverse temperature~$T$ of the environment in natural units, $F(\bcp) \equiv -\kT\ln \sum_x \exp\{-\beta E(x,\bcp) \}$ is the free energy, and $E(x,\bcp)$ is the system energy as a function of the microstate $x$ and a collection of experimentally controllable parameters $\bcp$ of the system. In the case of a gas confined to a cylinder a control parameter could be the position of the piston imposing a particular accessible volume. For a single macromolecule stretched between two optical traps, the control parameter could be the distance between the traps, imposing a harmonic energetic bias on the separation of the optical beads, between which the macromolecule is stretched. Control parameters can also represent collective variables, order parameters, or reaction coordinates.

In equilibrium, the macroscopic state of the system (the probability distribution over microstates) is completely specified by values of the control parameters, but out of equilibrium the system's probability distribution over microstates fundamentally depends on the history of the control parameter $\bcp$, which we denote by the control parameter \emph{protocol}~$\prot$. We assume the protocol is sufficiently smooth to be twice time-differentiable. 

\das{As formulated this driving by a time-dependent Hamiltonian can model a nonequilibrium steady state in the rest frame of a constantly-translating potential~\cite{Mazonka1999} or a driven damped harmonic oscillator that settles into a limit cycle. However it cannot treat a steady state, defined by time-independent probabilities for all system microstates in the lab reference frame. Such a steady state, for example induced by planar shear, is more naturally modeled by including a nonconservative force and corresponding dissipative flux~\cite{Williams:2008ft}.}

We adopt natural definitions of heat and work~\cite{Crooks:1998ue,Imparato:2007fd}: In accordance with the first law the average instantaneous rate of energy flow into the system, 
$\md \langle E \rangle_{\prot} / \md t$, 
is the sum of the average instantaneous rate of heat flow $\heat$ from the environment into the system, 
\be
\f{\md \langle \heat \rangle_{\prot}}{\md t} \equiv \l\langle \f{\md x^T}{\md t} \cdot \f{\pa E}{\pa x} \r\rangle_{\prot} \ ,
\ee
and the average instantaneous power $\power$ (rate of work $\work$) due to the external perturbation of the control parameters $\bcp$,
\be
\power \equiv \f{\md \langle \work \rangle_{\prot}}{\md t} \equiv - \f{\md\bcp^T}{\md t} \cdot \langle \fConj \rangle_{\prot} \ .
\ee
Here $\fConj \equiv - \pa E / \pa \bcp$ is the vector of forces conjugate to the control parameters $\bcp$, and angular brackets with subscript $\prot$ denote a nonequilibrium average over the ensemble following the control parameter protocol $\prot$. For the macromolecule with the trap position as the control parameter, the corresponding conjugate force would be the tension with which extension is resisted by the entire construct of a macromolecule, attached handles, and optical beads~\cite{Feng:2009vf}. 

For a system initially at equilibrium at time $\tCurr$ and control parameters $\bcpCurr$, the average heat flow vanishes, and the average power is 
\be
\power(\tCurr) = - \l[\f{\md\bcp^T}{\md t}\r]_{\tCurr} \cdot \langle \fConj \rangle_{\bcpCurr} \ ,
\ee
where the angled brackets with subscript $\bcpCurr$ denote an ensemble average at fixed control parameters $\bcpCurr$. 

Thus for a system at time $\tCurr$, following an arbitrary nonequilibrium protocol $\prot$ with current control parameter values $\bcpCurr$, the average \emph{excess} power exerted by the external agent on the system, over and above the average power on a system initially at equilibrium, is
\be
\instD = - \l[\f{\md\bcp^T}{\md t}\r]_{\tCurr} \cdot \langle \Delta \fConj(\tCurr) \rangle_{\prot} \ , \label{equ:exP}
\ee
where $\Delta \fConj (\tCurr) \equiv \fConj(\tCurr) - \la \fConj \ra_{\bcpCurr}$ is the deviation of the conjugate forces $\fConj$ at time $\tCurr$ from the average conjugate forces $\fConj$ at equilibrium under control parameter $\bcpCurr$. 

To first order in the magnitude of the external perturbation, \das{dynamic} linear-response theory expresses the deviation of the conjugate forces from equilibrium as an integrated response to the perturbation~\das{\cite{ZwanzigNESM}}, 
\be
\das{\langle \Delta \fConj(\tCurr) \rangle_{\prot}} \simeq \int_{-\infty}^{\tCurr} \md t' \ \bResp(\tCurr-t') \cdot [\bcp(t') - \bcpCurr] \ , \label{equ:LR}
\ee
where $\resp_{ij}(t) \equiv \beta \ \md\covarCurr_{ij}(t) / \md t$ represents the response of conjugate force $X_j$ at time $t$ to a perturbation in control parameter $\cp^i$ at time zero, and 
\be
\covarCurr_{ij}(t) \equiv \langle \delta X_j(0) \, \delta X_i(t) \rangle_{\bcp({\tCurr})}
\ee 
is the coefficient of the covariance matrix $\bCovarCurr(t)$ for conjugate forces $X_j$ and $X_i$ separated by time $t$, at constant control parameters $\bcpCurr$. 
\das{This approximation assumes that, over timescales where the response function $\md\bCovarCurr(t) / \md t \big|_{\tau}$ is significantly different from zero, both the nonequilibrium response $\langle \Delta \fConj(\tCurr) \rangle_{\prot}$ and the equilibrium change $\langle \fConj \rangle_{\bcp(t_0)} - \langle \fConj \rangle_{\bcp(t_0-\tau)}$ are linear in the control parameter change $\bcp(t_0)-\bcp(t_0-\tau)$. 

Substituting Eq.~\eqref{equ:LR} into Eq.~\eqref{equ:exP}} simplifies the nonequilibrium expectation:
\begin{align}
\instD = \beta \bigg[\f{\md\bcp^T}{\md t}&\bigg]_{\tCurr} \\
\cdot \int_{-\infty}^{\tCurr} \md t' &\f{\md \bCovarCurr(\tCurr-t')}{\md t'} \cdot [\bcpCurr - \bcp(t')] \ .\notag
\end{align}
Integration by parts gives
\be
\instD = \beta \l[\f{\md\bcp^T}{\md t} \r]_{\tCurr} \cdot \int_{-\infty}^{\tCurr} \md t' \, \bCovarCurr(\tCurr-t') \cdot \l[\f{\md \bcp}{\md t}\r]_{t'} ,
\ee
where the boundary term at $t' = \tCurr$ vanishes trivially, and the one at $t' = -\infty$ vanishes given that 
$
\lim_{t' \rightarrow -\infty} \bCovarCurr(t_0-t') \cdot \bcp(t') = {\bf 0} 
$.
This is satisfied, for example, when the system is initially at equilibrium.

We change integration variables to $\tDummy \equiv \tCurr - t'$ and Taylor expand the control parameter velocities at time $\tCurr-\tDummy$ around their values at $\tCurr$, 
\be
\l[ \f{\md\bcp}{\md t} \r]_{\tCurr-\tDummy} = \l[ \f{\md\bcp}{\md t} \r]_{\tCurr} - \ O\l( \l[ \f{\md^2\bcp}{\md t^2} \r]_{\tCurr} \r) \ .
\ee
When the control parameter velocities change on timescales slower than the relaxation time of the system's force fluctuations, we can keep only the constant term, yielding
\be
\instD = \l[ \f{\md\bcp^T}{\md t} \r]_{\tCurr}\!\! \cdot  \beta \int_0^{\infty}\md \tDummy  \bCovarCurr(\tDummy)  \cdot  \l[ \f{\md\bcp}{\md t} \r]_{\tCurr}  .
\ee

This analysis collapses the integral of the time-dependent covariance matrix into a single time-independent (equilibrium) matrix $\bfric\big(\bcpCurr\big)$ with entries
\begin{subequations}
\begin{align}
\fric_{ij}\Big(\bcpCurr\Big) &= \beta \int_0^{\infty} \md \tDummy \, \covarCurr_{ij}(\tDummy) \\
&= \beta \int_0^{\infty} \md \tDummy \, \langle \delta X_j(0) \, \delta X_i(\tDummy) \rangle_{\bcpCurr} \ .
\end{align}
\end{subequations}
This time-integrated force covariance matrix is the Kirkwood formulation of the friction tensor~\cite{Kirkwood:1946tq,Zwanzig:1964uc}.

We arrive at our central relations. The nonequilibrium excess power $\exPower$ along the protocol is determined by the friction tensor and experimentally controlled parameter velocities, 
\be
\instD = \l[ \f{\md\bcp^T}{\md t} \r]_{\tCurr} \cdot \bfric \big(\bcpCurr\big) \cdot \l[ \f{\md\bcp}{\md t} \r]_{\tCurr} \ .
\label{equ:centralResult}
\ee
This expression is entirely local: although in general nonequilibrium properties depend on the perturbation history, our relation only depends on the instantaneous values of the control parameter and its derivative.

The integral of the excess power over the control parameter protocol gives the mean excess work, $\exWork~=~\int_0^{\Delta t} \md t \, \exPower(t)$. This is the difference between the work the external agent does on the system during the nonequilibrium protocol and the work that would have been done if the protocol had been performed quasistatically, so that the system remained at equilibrium throughout. It follows from Eq.~\eqref{equ:centralResult} that the excess power scales as $|\md\bcp/\md t|^2$ and, for a given total transition time $\Delta t$, the excess work scales as $|\md\bcp/\md t|$. 

In general, at the conclusion of the protocol some fraction of this excess work will have been dissipated as heat to the environment, and the remaining fraction remains in the system, as an excess energy associated with being instantaneously out of equilibrium~\cite{Sivak:2012ab}. If at the conclusion of the protocol the system is allowed to fully equilibrate, all this remaining energy will be dissipated, and thus the excess work will equal the dissipation. 

Any covariance matrix is symmetric and positive-semidefinite~\cite{covPSDMatrix}, and the conjugate-force covariance matrix $\bCovarCurr(t)$ varies smoothly with control parameter values (except at macroscopic phase transitions). It follows that the friction tensor $\bfric\big(\bcp(\tCurr)\big)$, the time-integral of $\bCovarCurr(t)$, is symmetric, positive-semidefinite, and smoothly varying except at macroscopic phase transitions. Therefore the friction tensor $\bfric$ induces a Riemannian manifold on the space of thermodynamic states~\cite{McCleary}. Furthermore, positive-semidefiniteness of the friction tensor $\bfric$ guarantees that  excess power and work are non-negative, consistent with the second law.

This metric endows protocols with a number of useful properties. It defines a generalized thermodynamic length, $\mcL = \int_0^{\Delta t} \md t \, \sqrt{\exPower(t)}$ and divergence ${\mathcal J} = \Delta t \int_0^{\Delta t} \md t \, \exPower(t)$. The excess work is proportional to the thermodynamic divergence along the protocol, ${\mathcal J} =  \Delta t\ \exWork$. For a fixed control parameter path, the corresponding  thermodynamic length is independent of the time interval $\Delta t$ and the relative control parameter velocities en route, and places a lower bound on the excess work, $\exWork~\ge~\mcL^2/\Delta t$. By the Cauchy-Schwarz inequality, this bound is only realized for a protocol with a constant excess power~\cite{Salamon:1983vc,Crooks:2007wy}. 

Under a Riemannian metric the shortest paths (and therefore in our case the optimal, minimum excess work protocols) are geodesics, the closest thing to a straight line in a curved manifold. This property should simplify the discovery of optimal protocols in complicated energy landscapes. Moreover, from the definition and scaling of the thermodynamic length and divergence, it follows that the control parameter path of an optimal protocol is independent of the protocol duration. Increasing or decreasing the duration does not change the optimal path in the linear-response regime. Finally, we note that the metric structure ensures that the excess work will be invariant to linear transformations of the control parameters.

\paragraph*{\bf Discussion.}
The present formalism generalizes several other frameworks in the existing literature on thermodynamic length. The control parameter friction tensor can be decomposed into 
\be
\bfric\big( \bcpCurr \big) = \kT \ \tRB\big(\bcpCurr\big)\circ\FIB\big(\bcpCurr\big) \ ,
\ee
the Hadamard product (entry-by-entry product) `$\circ$' of the integral relaxation time matrix $\tRB$ and the Fisher information matrix $\FIB$, scaled by $\kT$. The Fisher information is defined as~\cite{coverThomas}:
\be
\FI_{ij}(\bcp) \equiv \l\langle \f{\pa \ln \pi(x|\bcp)}{\pa\cp^i} \f{\pa \ln \pi(x|\bcp)}{\pa\cp^j} \r\rangle_{\bcp} \ .
\ee
For a system in thermal equilibrium \eqref{equ:CanonEns}, this simplifies to $\beta^2\, \l\langle \delta X_j \, \delta X_i\r\rangle_{\bcp}$, the covariance of the forces conjugate to control parameters $\cp^i$ and $\cp^j$~\cite{Crooks:2007wy}. The integral relaxation time~\cite{Garanin:1990tg},
\be
\tR_{ij}\l(\bcp\r) \equiv \int_0^{\infty} \md \tDummy \, \f{\langle \delta X_j(0)\, \delta X_i(\tDummy) \rangle_{\bcp}}{\langle \delta X_j\, \delta X_i \rangle_{\bcp}} \ ,
\ee
generalizes to multiple dimensions and nonexponential relaxation kinetics the more familiar relaxation time $\tSimple$ (the time constant in exponential relaxation kinetics  where correlations decay over time according to $\exp[-t/\tSimple]$).

Equation~\eqref{equ:centralResult} allows for an integral relaxation time that varies with the current control parameter value. When the relaxation time does not vary with the control parameter, the Riemannian metric reduces to the Fisher information metric~\cite{Rao1945}, recovering the microscopic thermodynamic length formulation of~\cite{Crooks:2007wy}. Nevertheless, the generalization to varying integral relaxation time is important, as in many systems we expect the relaxation time to vary substantially, especially near transition interfaces separating metastable basins of attraction. 

In general the Fisher information is related to Schl\"ogl's metric, the second derivative of the free entropy $\psi \equiv -\beta F$, by $\FIB(\bcp) = -\f{\pa^2\beta F}{\pa\bcp^2} + \l\langle \f{\pa^2\beta E}{\pa\bcp^2} \r\rangle$. The last term vanishes when the energy is linear in the control parameters, when the control parameters are intensive, or when the control parameters are extensive in the macroscopic limit. In these limits, we recover the original formalism as the differential geometry of thermodynamic potentials~\cite{Weinhold:1975wr,Ruppeiner:1979ve,Schlogl1985,Salamon:1984wv}.

Expressing the metric in terms of a friction tensor in thermodynamic state space helps clarify the nature of thermodynamic length. Roughly speaking, this friction represents the resistance of the system to control parameter changes that are imposed in finite time. According to Eq.~\eqref{equ:centralResult}, for a fixed control parameter velocity $\md \cp/\md t$, the excess power is greater where the control parameter friction coefficient $\fric$ is greater.  Hence the excess work is reduced when the protocol proceeds slower in regions of a high friction coefficient, which can occur when the relaxation time or equilibrium fluctuations are large.

In Cartesian space, the friction tensor (also known as the inverse diffusion tensor) has been posited as a metric tensor~\cite{Fris1966,Haenggi:1982tm,Ottinger1997}, and used to understand the connectivity of white matter in the brain~\cite{ODonnell2002} and the paths of electrical excitation waves in cardiac tissue~\cite{Wellner:2002hb}.  de Koning and Antonelli~\cite{deKoning:1997vv} derived a similar expression for excess work (but assumed simple exponential relaxation time) using similar linear-response arguments. Tsao, \emph{et al.}~\cite{Tsao:1994wc} derived a similar expression assuming endoreversibility (but with the integral relaxation time replaced by the endoreversible \emph{lag time}, the elapsed time since the control parameters had values for which the current distribution over microstates is an equilibrium distribution). The case of nonexponential kinetics, the connections to thermodynamic length analysis, and the interpretation in terms of a friction tensor in thermodynamic state space have not to our knowledge been previously reported.

\paragraph*{\bf Applications.}
We now demonstrate a simple expression for the control parameter protocol that minimizes the excess work for a system with one control parameter, given that it must transition between points $\cp_a$ and $\cp_b$ in fixed time $\Delta t$.  The optimal control parameter protocol is found via the Euler-Lagrange equation~\cite{morin} for this problem, with cost function $f(\cp(t),\dot{\cp}) = \fric(\cp)\ \dot{\cp}^2$,  
\be
0 = \f{\pa f}{\pa \cp} - \f{\md}{\md t}\l[ \f{\pa f}{\pa\dot{\cp}} \r] = -2\fric(\cp)\ \ddot{\cp} - \fric'(\cp)\ \dot{\cp}^2 \ .
\ee
This has the solution
\be
\dot{\cp}^{\text{opt}}(t)  =  \f{ (\cp_b-\cp_a)\  \fric\big(\cp(t)\big) ^{-\frac{1}{2}}}{\int_0^{\Delta t} \md t'  \ \fric\big(\cp(t')\big) ^{-\frac{1}{2}}} \  \propto\  \fric\big(\cp(t)\big) ^{-\frac{1}{2}} \label{equ:optProtOne} \ .
\ee
Thus, in the single control parameter case, the optimal protocol proceeds with a velocity inversely proportional to the square root of the control parameter friction, evaluated for the \emph{current} value of the control parameter, and hence [using Eq.~\eqref{equ:centralResult}] has a constant excess power. Where the system experiences large friction in thermodynamic state space, the optimal control parameter protocol will change slowly. The time interval $\Delta t$ only sets the proportionality constant, but not the relative velocities at different points in the protocol. 

The general multiple-parameter case does not admit of a straightforward optimization. But we can extend our analysis to the optimization of a particular two-parameter protocol: a particle diffusing in a harmonic potential according to the overdamped Langevin equation with Cartesian friction coefficient~$\fricSp$, with control parameters the location $\springMin$ and spring constant $k$ of the harmonic potential. We perturb the system from $\big(\springMin_a,k_{a} \big)$ to $\big( \springMin_b,k_{b} \big)$ in a fixed time interval $\Delta t$. The control parameter friction tensor is
\be
\bm \fric = 
\left( \begin{array}{cc}
\fricSp & 0 \\
0 & \f{\fricSp}{4\beta k^3} \end{array} \right)\ .
\ee
Since this matrix is diagonal, and the $k,k$ term is independent of $y$ and the $y,y$ term is independent of $k$, the protocol can be optimized for each control parameter separately using Eq.~(\ref{equ:optProtOne}), yielding for the optimal protocol
\begin{align}
\f{\md}{\md t} \, \springMin  =  \f{\springMin_b - \springMin_a}{\Delta t} \ ,  \quad
\f{\md }{\md t} \, k^{-1/2}  =  \f{k_b^{-1/2} - k_a^{-1/2}}{\Delta t} \label{equ:optimalK}\ .
\end{align}
Under the optimal protocol the trap center and inverse square root of the spring constant each change at a constant rate, independent of time. This corresponds to changing the equilibrium mean and standard deviation of position at a constant rate.

Seifert and coworkers elegantly derived the exact optimal protocols for perturbing the position and spring constant separately, for both over-damped~\cite{Schmiedl:2007ei} and under-damped~\cite{GomezMarin:2008jz} Langevin dynamics. Their analysis found optimal protocols similar to ours but with discrete control parameter jumps at the beginning and end of the protocol \das{(though these jumps are smoothed to boundary layers under regularization that penalizes acceleration~\cite{Aurell:2012fz})}. Our method misses such protocol jumps because our derivation assumes that the velocities of protocols change smoothly. Other than the discrete jumps at the boundaries, our approximation produces results that near equilibrium differ from optimal values by, to leading order, a dimensionless measure of distance from equilibrium, $\fricSp/(k\Delta t)$. \das{Such exact results are useful where they are tractable, but the thermodynamic metric provides a convenient, general computational framework, especially in complex systems.}

\paragraph*{\bf Acknowledgments.}
The authors thank John~D.~Chodera (University of California, Berkeley) for enlightening discussions and constructive feedback on the manuscript. This work was supported by the Director, Office of Science, Office of Basic Energy Sciences, of the U.S. Department of Energy under Contract No. DE-AC02-05CH11231.

%

\end{document}